\documentclass[aps,pre,reprint,amsmath,amssymb,superscriptaddress]{revtex4-2}

\usepackage{graphicx}
\usepackage{bm}
\usepackage{xcolor}
\usepackage[colorlinks=true,linkcolor=blue,citecolor=blue,urlcolor=blue]{hyperref}

\begin{document}

\title{\Large Thermodynamic Crossovers in Supercritical Matter: \\One, No One, and One Hundred Thousand}
\author{Yang Xu}
\author{Matteo Baggioli}
\email{b.matteo@sjtu.edu.cn}
\affiliation{School of Physics and Astronomy, Shanghai Jiao Tong University, Shanghai 200240, China}
\affiliation{Wilczek Quantum Center, School of Physics and Astronomy, Shanghai Jiao Tong University, Shanghai 200240, China}

\begin{abstract}
Beyond a critical point, two thermodynamic phases become continuously connected, without being separated by any thermodynamic singularity. The paradigmatic example is the liquid–gas transition, where several criteria have been proposed to distinguish liquid-like and gas-like regimes within the supercritical region. Recently, a pair of thermodynamic crossover lines, $L^\pm$, was introduced to address a proposed incompatibility of previous definitions with the universal critical behavior of the Ising model. These lines divide the supercritical region into three regimes, meet only at the critical point, and obey the expected critical scaling laws. Here, we demonstrate that these requirements do not uniquely determine the crossover boundaries. Instead, infinitely many pairs of lines satisfy the same conditions, with different choices of thermodynamic paths selecting different crossover boundaries. We further show that the Lee–Yang formalism, despite providing an intrinsic characterization of critical singularities, does not resolve this ambiguity in defining real-axis crossover boundaries. Restricting to regular probing paths that respect the $Z_2$ symmetry of the critical point recovers the Widom and $L^\pm$ prescriptions asymptotically. This restriction, however, is a choice of protocol rather than a property of the system, and even then two inequivalent prescriptions survive. Our results demonstrate that universal critical behavior alone does not select privileged supercritical thermodynamic crossover boundaries. While such lines may identify physically meaningful changes in thermodynamic, structural, or transport properties, their precise locations and the widths of the regions they delimit remain dependent on the operational criteria used to define them.
\end{abstract}

\maketitle

\section{Introduction}
A thermodynamic phase transition corresponds to a nonanalyticity of the thermodynamic potential, whose location is independent of the path chosen to approach it \cite{Landau1980Statistical}. However, two phases sharing the same symmetry need not be separated by a thermodynamic singularity under all conditions. The paradigmatic example is the liquid–gas transition, whose coexistence line terminates at a critical point, beyond which the two phases can be continuously connected without encountering any genuine phase transition \cite{kiran2012supercritical,proctor2020liquid}.

More precisely, beyond the critical point, in the so-called supercritical regime, the distinction between liquid and gas ceases to be thermodynamically sharp, challenging their identification as two distinct phases of matter \cite{10.1063/PT.3.1796}. For water, this occurs above the critical temperature $T_c \approx 647\,\mathrm{K}$ and critical pressure $P_c \approx 22\,\mathrm{MPa}$.

The disappearance of the thermodynamic boundary between liquid and gas was first observed in the nineteenth century, through the pioneering experiments of Charles Cagniard de la Tour \cite{de1822expose} and the subsequent systematic studies of Thomas Andrews \cite{Andrews1869}. More than a century later, considerable effort has been devoted to reconstructing boundaries within this same region, using disparate criteria ranging from structural properties (e.g., the Fisher--Widom line \cite{fisher1969decay}) and dynamical behavior (e.g., the Frenkel line \cite{PhysRevE.85.031203}) to thermodynamic response functions (e.g., the Widom line \cite{xu2005relation}). Despite the absence of genuine thermodynamic phase transitions, these constructions aim to distinguish liquid-like and gas-like regimes within supercritical matter. In this work, we focus specifically on thermodynamic criteria and examine to what extent the resulting crossover boundaries represent intrinsic properties of the system.

Undoubtedly, the most widely studied thermodynamic crossover is the Widom line \cite{xu2005relation}, commonly defined as the locus of maxima of a thermodynamic response function (e.g., heat capacity or compressibility) along prescribed thermodynamic paths, such as isobars or isotherms. By construction, the Widom line emanates from the critical point, where the corresponding response functions exhibit the expected universal critical scaling. Despite substantial experimental evidence supporting its physical relevance \cite{Simeoni2010,Maxim2019}, its definition depends on the response function considered. Moreover, sufficiently far from the critical point, the corresponding maxima disappear, rendering the construction inapplicable.

More recently, it has been argued that the Widom line, together with other proposed supercritical crossovers, is incompatible with the universal critical behavior of the Ising model \cite{li2024thermodynamic}. To address this proposed inconsistency, two novel thermodynamic crossover lines, $L^\pm$, were introduced as the loci of maxima of a thermodynamic response function evaluated along paths parallel to the critical isochore. While this construction remains dependent on the chosen response function and likewise ceases to be applicable far from criticality, it leads to a substantially different physical picture. In particular, the two lines delimit three supercritical regimes: liquid-like, gas-like, and an intermediate region in which the two become indistinguishable, rather than simply forming a mixture of the former two. The proposed $L^\pm$ lines were motivated by three requirements: (i) they meet only at the critical point, (ii) they obey the expected universal critical scaling laws, and (iii) they separate the supercritical region into three distinct regimes.

The $L^\pm$ construction has since been extended to a broad variety of systems, ranging from classical statistical models to quantum and gravitational settings \cite{Lv2025,hgwk-pzw2,doi:10.34133/research.1406}. More recently, distinct structural, transport, and dynamical signatures associated with crossing these lines have been identified \cite{3mxm-gqh5}. Nevertheless, whether these observations genuinely support the existence of three distinct supercritical states, or can instead be understood as different manifestations of a continuous crossover between liquid-like and gas-like behavior, remains an open question.

Here, we critically revisit the physical meaning of these thermodynamic crossover lines and examine whether they can be interpreted as privileged boundaries between distinct supercritical states. In particular, we address a fundamental question: if infinitely many crossover lines satisfy the same universal requirements, what distinguishes one particular construction from the others, and on what grounds can it be regarded as an intrinsic property of matter? We also identify the additional requirement that singles out the Widom and $L^\pm$ constructions, show that it constrains the probing protocol rather than the system, and discuss its status in real fluids.
\section{Infinitely many crossover lines}
\label{sec:infinite}
Close to a critical point, the singular part of the free-energy density takes the generalized homogeneous form
\begin{equation}
    f_s(t,h)=\ell^{-d}\,\Psi\left(\ell^{y_t}t,\ell^{y_h}h\right),
\end{equation}
where $d$ is the spatial dimension and $\ell$ is the renormalization-group (RG) rescaling factor. For simplicity, we retain only two relevant thermodynamic scaling fields: the reduced temperature-like variable $t$, measuring the distance from the critical temperature, and the ordering field $h$, conjugate to the order parameter $m$. Their corresponding RG eigenvalues are $y_t$ and $y_h$, respectively. For the Ising model, $t=(T-T_c)/T_c$ and $h$ is the magnetic field (in suitable units). In liquid--gas systems, these variables correspond more generally to suitable combinations of the physical thermodynamic fields near the critical point.

Choosing $\ell=|t|^{-1/y_t}$, the homogeneous form becomes
\begin{equation}
    f_s(t,h)=|t|^{d/y_t}\,
    \mathcal{F}_{\pm}(x),
    \qquad
    x\equiv\frac{h}{|t|^{y_h/y_t}},
\end{equation}
where the subscripts $\pm$ refer to the regions above and below the critical temperature. Using the standard scaling relations, this can be expressed in terms of the critical exponents as
\begin{equation}
    f_s(t,h)=|t|^{2-\alpha}\,
    \mathcal{F}_{\pm}(x),
    \qquad
    x\equiv\frac{h}{|t|^\Delta},
    \qquad
    \Delta\equiv\beta\delta.
\end{equation}
Here, $\alpha$, $\beta$, and $\delta$ are defined by
\begin{equation}
    C\propto |t|^{-\alpha},
    \qquad
    m\propto(-t)^\beta,
    \qquad
    m\propto \operatorname{sgn}(h)|h|^{1/\delta},
\end{equation}
where $C$ denotes the singular heat capacity, while the latter two relations describe the spontaneous order parameter for $t<0$ at zero ordering field and the critical isotherm $t=0$, respectively.

In the following, we focus exclusively on the supercritical region $t>0$ and omit the $+$ subscript from the scaling function.

The order parameter $m$ and the susceptibility $\chi$ follow directly from the free energy,
\begin{equation}
    m\equiv-\frac{\partial f_s}{\partial h}
    =t^\beta\mathcal{M}(x),
    \qquad
    \chi\equiv\frac{\partial m}{\partial h}
    =t^{-\gamma}\mathcal{X}(x).
\end{equation}
More generally, a thermodynamic response function governed by a homogeneous scaling law can be written as
\begin{equation}
    R(t,h)=t^{-\kappa}\mathcal{R}(x).
    \label{r1}
\end{equation}

Consider now a family of thermodynamic paths defined by
\begin{equation}
    q(t,h)=\mathrm{const.}
\end{equation}
The stationary points of $R$ along these paths satisfy
\begin{equation}
    \left.\frac{dR}{dt}\right|_q=0.
\end{equation}
Using the scaling form in Eq.~\eqref{r1}, this condition becomes
\begin{equation}
    \kappa\mathcal{R}(x)
    +
    \left[
    \Delta x-t^{1-\Delta}
    \left.\frac{dh}{dt}\right|_q
    \right]\mathcal{R}'(x)=0.
    \label{ex}
\end{equation}
This expression already reveals the origin of the ambiguity discussed below: the location of a thermodynamic response extremum depends not only on the scaling function but also on the direction along which the response is probed, $\left.\frac{dh}{dt}\right|_q$.

To construct an explicit set of crossover lines compatible with universal critical scaling, we consider families of paths of the form
\begin{equation}
    q(t,h)=h-\zeta t^\Delta=\mathrm{const.},
    \label{path}
\end{equation}
where $\zeta$ parametrizes the choice of thermodynamic paths. This form is motivated by the fact that $h$ and $t^\Delta$ have the same RG scaling dimension, so that $q$ transforms as an ordering field, $q\rightarrow \ell^{y_h}q$. As shown below, this choice also renders the extremum condition independent of the distance from the critical point when expressed in terms of the scaling variable $x$. Importantly, Eq.~\eqref{path} represents only a restricted family of scaling-compatible paths, which is nevertheless sufficient to demonstrate the nonuniqueness of crossover boundaries.

For noninteger $\Delta$, the paths of Eq.~\eqref{path} are nonanalytic at $t=0$. They are nevertheless smooth and analytic throughout the supercritical region $t>0$. Moreover, the extremum condition, Eq.~\eqref{ex}, depends only on the local tangent $\left.dh/dt\right|_q$ at the extremum, so the behavior of the paths away from the maxima plays no role. Their nonanalytic behavior at $t=0$ does not imply any additional thermodynamic singularity or invalidate their use as trajectories along which response-function maxima are evaluated.

The specific form of Eq.~\eqref{path} is, however, not essential. Simple
power-law paths, $h=Q\,t^{p}$, also yield crossover lines
$h_\pm\propto t^\Delta$ whenever off-axis maxima exist: the choice $p=0$
reproduces the $L^\pm$ lines, whereas, e.g., $p=-1$ gives a different
symmetric pair (see Sec.~\ref{sec:special}).

Along these paths,
\begin{equation}
    \left.\frac{dh}{dt}\right|_q
    =\zeta\Delta t^{\Delta-1},
\end{equation}
and Eq.~\eqref{ex} reduces to
\begin{equation}
    \kappa\mathcal{R}(x)
    +\Delta(x-\zeta)\mathcal{R}'(x)=0.
    \label{extremum_scaled}
\end{equation}
Crucially, this equation depends only on the scaling variable $x$, with no explicit dependence on $t$. For a fixed value of $\zeta$, let us denote by $c_+(\zeta)>0$ and $c_-(\zeta)<0$ two solutions corresponding to maxima of the response function on the positive- and negative-field sides, respectively. Their loci in the thermodynamic plane are then
\begin{equation}
    L_\zeta^\pm:\qquad
    h_\pm(t)=c_\pm(\zeta)t^\Delta.
    \label{fam}
\end{equation}
These curves are obtained by collecting the positions of the maxima along the paths defined by Eq.~\eqref{path}, while varying the constant $q$ at fixed $\zeta$. Thus, the paths along which the extrema are evaluated must be distinguished from the resulting crossover lines.

For every admissible value of $\zeta$, Eq.~\eqref{fam} defines a pair of crossover lines that approach the critical point as $t\rightarrow0^+$ and divide the supercritical scaling region into three sectors. By construction, all these lines satisfy
\begin{equation}
    h_\pm\propto t^{\beta\delta},
    \qquad
    m_\pm=t^\beta\mathcal{M}(c_\pm),
    \label{amp}
\end{equation}
and therefore reproduce the universal critical exponents required in Ref.~\cite{li2024thermodynamic}, provided the corresponding order-parameter amplitudes are nonzero.

The relation between the choice of thermodynamic paths and the resulting crossover amplitudes can be made explicit by solving Eq.~\eqref{extremum_scaled} for $\zeta$,
\begin{equation}  
    \zeta(c)=c+
    \frac{\kappa\mathcal{R}(c)}
    {\Delta\mathcal{R}'(c)}.
    \label{path_amplitude}
\end{equation}
This establishes a direct correspondence between the parameter specifying the thermodynamic paths and the location of the corresponding extrema. Within any interval where the solutions are nondegenerate maxima and this relation is invertible, distinct choices of $\zeta$ generate distinct crossover lines. Since $\zeta$ can vary continuously, infinitely many such lines can be constructed from the same thermodynamic response function, all satisfying the same critical scaling laws.

We use the same paths, with the same value of $\zeta$, to find both branches. This way, each crossover definition follows one clear rule. For a response function symmetric under $h\rightarrow-h$, if $c$ is a solution of Eq.~\eqref{extremum_scaled} for $\zeta$, then $-c$ is a solution for $-\zeta$, so that the two branches are related by $c_-(\zeta)=-c_+(-\zeta)$. For $\zeta\neq0$, the resulting crossover lines are therefore not symmetric under $h\rightarrow-h$. As discussed in Sec.~\ref{sec:special}, this reflects the fact that, for $\zeta\neq0$, the paths of Eq.~\eqref{path} are not themselves invariant under $h\rightarrow-h$.

Two familiar crossover constructions emerge as particular choices of thermodynamic paths. First, consider the conventional Widom prescription \cite{xu2005relation}, in which the response function is maximized along paths of constant temperature, $t=\mathrm{const.}$ For an Ising-symmetric susceptibility, the maximum occurs at $x=0$, yielding a single crossover line $h=0$. In this case, the positive- and negative-field boundaries collapse onto the same line, $L^+=L^-$, and no intermediate region of finite width is identified.

Second, choosing $\zeta=0$ in Eq.~\eqref{path} corresponds to paths of constant ordering field, $h=\mathrm{const.}$ The corresponding extremum condition becomes
\begin{equation}
    \kappa\mathcal{R}(x)
    +\Delta x\mathcal{R}'(x)=0,
    \label{ext0}
\end{equation}
which, for the susceptibility, reproduces the two distinct $L^\pm$ crossover lines proposed in Ref.~\cite{li2024thermodynamic}. These constructions therefore arise from the same underlying thermodynamic response function but employ different directions along which its maxima are identified.

This comparison also clarifies an important conceptual point. In Ref.~\cite{li2024thermodynamic}, the conventional single-line Widom construction was argued to be incompatible with the critical universality of the Ising model, on the grounds that the latter requires two distinct crossover lines. Our analysis shows that this conclusion does not follow from universal scaling theory. Both the single-line Widom prescription and the two-line $L^\pm$ construction can be obtained from the same underlying scaling function, by evaluating the maxima of the same susceptibility along different thermodynamic paths. The number and location of the resulting crossover lines are therefore not dictated by the Ising universality class alone, but depend on the operational prescription used to identify them. This is fundamentally different from genuine thermodynamic phase boundaries identified as singularities in the thermodynamic potential. It simply reflects that, within the supercritical region, the underlying thermodynamics selects no unique crossover line: there are only continuous crossovers, which cannot be represented by lines without introducing unavoidable ambiguities.

Indeed, the $Z_2$ symmetry of the Ising model is compatible both with a single invariant crossover line at $h=0$ and with a pair of symmetry-related lines at $h_\pm=\pm c\,t^\Delta$. The former is mapped onto itself under $h\rightarrow-h$, whereas the latter two lines are exchanged. Thus, symmetry does not require two distinct crossover boundaries. In particular, the requirement that two lines emerge from the critical point is an additional criterion, not a consequence of critical universality. More generally, our construction shows that infinitely many pairs satisfy the same critical scaling laws, further demonstrating that universality alone does not select a privileged partition of the supercritical region.

For the same reason, the Widom line is not in conflict with critical scaling. The line $h=0$ corresponds to $x=0$ and therefore belongs to the same family of RG-invariant curves, $x=\mathrm{const.}$, as the $L^\pm$ lines. Response functions evaluated along it obey the universal scaling laws, e.g., $\chi\propto t^{-\gamma}$ and $C\propto t^{-\alpha}$. The only property it lacks is a nonzero order-parameter amplitude in Eq.~\eqref{amp}, since $m$ vanishes identically on $h=0$ by symmetry. In liquid--gas systems, Widom lines likewise approach the $h=0$ direction at the critical point. It is therefore the requirement of a nonzero amplitude, $m_\pm\propto t^\beta$, motivated by analogy with the subcritical coexistence curve, $m\propto(-t)^\beta$, that excludes the Widom line, rather than critical scaling itself. This analogy is not enforced by any symmetry relating $t$ and $-t$. Moreover, it is satisfied by every member of the family in Eq.~\eqref{fam}, so it does not single out the $L^\pm$ lines either.

Importantly, the choice of thermodynamic paths determines not only the location of the two supercritical crossover lines but also the width of the proposed intermediate indistinguishable regime. This width is simply given by
\begin{equation}
    \mathcal{W}(t,\zeta)
    =h_+(t)-h_-(t)
    =\left[c_+(\zeta)-c_-(\zeta)\right]t^\Delta.
\end{equation}
By construction, the width follows the expected critical scaling, $\mathcal{W}(t,\zeta)\propto t^{\beta\delta}$, but its amplitude depends on the operational definition adopted. Consequently, the extent of the proposed intermediate regime can be continuously adjusted by changing the crossover prescription, without modifying the underlying thermodynamics. While this does not exclude the existence of physically meaningful crossovers, it demonstrates that the width of a region defined solely through response-function maxima cannot, by itself, be regarded as an intrinsic property of matter.

As shown explicitly for the Landau model in Sec.~\ref{ggl}, each amplitude $c_\pm(\zeta)$ varies monotonically with $\zeta$, so that each branch individually sweeps the entire supercritical scaling region on its side, from the Widom line ($c\to0$) to the critical isotherm ($|c|\to\infty$), and distinct lines of the same branch never cross. The $L^\pm$ lines are therefore interior members of this family, rather than its envelope or its outer bound, and $\mathcal{W}$ can be made arbitrarily large at any fixed distance from the critical point.

It is instructive to ask which crossover lines are obtained from generic, analytic families of thermodynamic paths. Since Eq.~\eqref{ex} depends only on the local tangent of the path, it is convenient to define the effective path parameter $\zeta_{\rm eff}(t,h)\equiv t^{1-\Delta}\left.(dh/dt)\right|_q/\Delta$, in terms of which the extremum condition takes the form of Eq.~\eqref{extremum_scaled} with $\zeta\to\zeta_{\rm eff}$. For an analytic family with nonzero slope $s_0=\left.dh/dt\right|_q$ at the critical point, $\zeta_{\rm eff}\sim s_0t^{1-\Delta}\to\infty$ because $\Delta>1$: one branch approaches the Widom line, while the other does not reach the critical point. Only when the paths are tangent to the $h=0$ direction at the critical point does $\zeta_{\rm eff}$ vanish, reproducing the $L^\pm$ lines irrespective of higher-order curvature. Intermediate asymptotic amplitudes, $0<|\zeta|<\infty$, require direction fields scaling as $t^{\Delta-1}$, as in Eq.~\eqref{path}, which are nonanalytic at $t=0$. This nonanalyticity is, however, confined to the limit $t\to0$: at any finite distance from the critical point, the same maxima are obtained along analytic paths sharing the same local tangents, so that the nonuniqueness established above holds throughout the region accessible to experiments and simulations. Whether tangency to the $h=0$ direction, and hence the $L^\pm$ construction, can be regarded as physically privileged is examined in Sec.~\ref{sec:special}.

\section{Sensitivity to the regular part of the thermodynamic potential}
\label{two}
An additional distinction between genuine critical points and supercritical crossover lines concerns their sensitivity to nonsingular contributions to the thermodynamic potential. A true critical point is characterized by a nonanalyticity of the free energy, whereas supercritical crossover lines are commonly defined through extrema of thermodynamic response functions. We now show that this difference has a direct consequence: while the location of a genuine critical singularity is unaffected by the addition of an analytic background, the locations of supercritical response maxima generally are not.

Consider the decomposition of the equilibrium free-energy density into singular and regular contributions,
\begin{equation}
    f(t,h)=f_{\mathrm s}(t,h)+f_{\mathrm{reg}}(t,h).
\end{equation}
Let us modify the free energy by adding an arbitrary analytic function $g(t,h)$,
\begin{equation}
    \widetilde f(t,h)=f(t,h)+\varepsilon g(t,h),
\end{equation}
where $\varepsilon$ controls the strength of the modification. Since $g(t,h)$ is analytic in a neighborhood of the critical point, this addition cannot remove or shift the nonanalyticity associated with the critical point. In particular, the position of the critical singularity, its critical exponents, and the leading singular scaling functions remain unchanged.

The situation is different for supercritical crossover lines. Consider, for concreteness, the susceptibility
\begin{equation}
    \chi(t,h)=-\frac{\partial^2f}{\partial h^2}.
\end{equation}
Under the analytic modification, it transforms as
\begin{equation}
    \widetilde\chi(t,h)
=
\chi(t,h)-\varepsilon\frac{\partial^2g}{\partial h^2}.
\end{equation}
Suppose that the crossover line is defined through maxima of $\chi$ along a prescribed family of thermodynamic paths $q(t,h)=\mathrm{const.}$ Introducing the derivative along these paths,
\begin{equation}
    D_q\equiv
\frac{\partial}{\partial t}
+
\left(\frac{dh}{dt}\right)_q
\frac{\partial}{\partial h},
\end{equation}
the original crossover condition reads
\begin{equation}
    D_q\chi=0.
\end{equation}
After adding the analytic contribution, the condition becomes
\begin{equation}
    D_q\chi
-\varepsilon D_q\frac{\partial^2g}{\partial h^2}
=0.
\end{equation}
Crucially, the second term is generally nonzero, even though the added contribution is completely nonsingular.

To make this dependence explicit, let $t_*(q)$ denote a nondegenerate maximum of the original susceptibility along a fixed path, satisfying
\begin{equation}
    D_q\chi\big|_{t_*}=0,
\qquad
D_q^2\chi\big|_{t_*}<0.
\end{equation}
For sufficiently small $\varepsilon$, the implicit function theorem gives the displacement of this maximum as
\begin{equation}
    \widetilde t_*(q)-t_*(q)
=
\varepsilon
\left.
\frac{D_q\partial_h^2g}
{D_q^2\chi}
\right|_{t_*}
+O(\varepsilon^2).
\end{equation}
Thus, even when the thermodynamic paths and the response function are kept fixed, the corresponding crossover maxima generally shift under the addition of an analytic contribution to the free energy. By contrast, the location of the genuine critical singularity remains exactly unchanged.

This establishes a second source of nonuniqueness, independent of the path dependence demonstrated in the preceding section. Thermodynamic crossover lines are sensitive not only to the operational procedure used to identify them, but also to nonsingular details of the thermodynamic potential that do not affect the underlying critical singularity.

Importantly, this sensitivity is compatible with universal critical scaling. For a divergent susceptibility, the singular contribution dominates sufficiently close to the critical point, so analytic backgrounds generally produce subleading corrections to the crossover locations without changing their leading critical exponents. Nevertheless, away from the asymptotic critical region, such corrections need not be negligible. Consequently, while critical singularities are protected against analytic modifications of the free energy, supercritical crossover lines possess no analogous protection.
\section{Concrete examples}
\subsection{Landau model}
\label{ggl}
Consider the Landau free-energy density
\begin{equation}
    f(m;t,h)=\frac{a}{2}t\,m^2+\frac{b}{4}m^4-hm,
\qquad a,b>0
\end{equation}
where $t$ is the reduced temperature, $h$ the external field and $m$ the order parameter (magnetization).

The equilibrium order parameter satisfies
\begin{equation}
    \frac{\partial f}{\partial m}=0,
\end{equation}
yielding the equation of state
\begin{equation}
    h=atm+bm^3.
\end{equation}
The susceptibility can be easily derived,
\begin{equation}
    \chi=\frac{\partial m}{\partial h}
=\frac{1}{at+3bm^2},
\end{equation}
and in absence of an external field, $h=0$, diverges at the critical point as $\chi \propto t^{-1}$.

It is convenient to introduce the following dimensionless scaling variables,
\begin{equation}
    m=\sqrt{\frac ab}\,t^{1/2}y,
\qquad
h=\frac{a^{3/2}}{\sqrt b}\,t^{3/2}x,
\end{equation}
in terms of which the equation of state and the susceptibility take the simple form
\begin{equation}
x=y+y^3, \qquad
\chi=\frac{1}{at}\frac{1}{1+3y^2}.
\end{equation}
The critical point at $t=h=0$ is characterized by the mean-field critical exponents
\begin{equation}
\beta=\frac12,\qquad
\gamma=1,\qquad
\delta=3,\qquad
\Delta=\frac32.
\end{equation}
Using the normalized ordering field
\begin{equation}
    H=\frac{\sqrt b}{a^{3/2}}h,\label{fff}
\end{equation}
we consider the paths
\begin{equation}
    q=H-\zeta t^{3/2}=\mathrm{const}.
\end{equation}
The general extremum condition, Eq.~\eqref{extremum_scaled}, becomes
\begin{equation}
   \mathcal{X}(x)+\frac32(x-\zeta)\mathcal{X}'(x)=0, 
\end{equation}
where
\begin{equation}
    \mathcal{X}(x)=\frac{1}{1+3y^2},
\qquad x=y+y^3.
\end{equation}
Because of the simplicity of this model, the extremum condition can be solved exactly
\begin{equation}
\zeta(y)=\frac y3-\frac{1}{9y}.
\end{equation}
Equivalently, solving for $y$,
\begin{equation}
y_\pm(\zeta)
=\frac{3\zeta\pm\sqrt{9\zeta^2+4/3}}{2}.\label{sol1}
\end{equation}
The corresponding crossover lines are therefore analytically given by
\begin{equation}
    L_\zeta^\pm:\quad
h_\pm(t)=\frac{a^{3/2}}{\sqrt b}
\left[y_\pm(\zeta)+y_\pm^3(\zeta)\right]t^{3/2}.
\end{equation}
Thus, we have explicitly constructed a continuous infinity of crossover-line pairs within the same mean-field theory.

Importantly, it is easy to verify that each of these solutions corresponds to a maximum of the susceptibility and that the relation between $\zeta$ and $y$ is strictly monotonic, $d \zeta/dy>0$. We have therefore established an exact one-to-one correspondence between each path parameter and the location of the two susceptibility maxima. Notice that, for each finite $\zeta$, $y_+(\zeta)>0,\, y_-(\zeta)<0$. Hence, both branches always exist and meet only at the critical point. They also divide the supercritical scaling plane into three regions, exactly as required in \cite{li2024thermodynamic}.

For fixed ordering field, $\zeta=0$, Eq.~\eqref{sol1} gives $y_\pm(\zeta)
=\pm 1/\sqrt{3}$ and,
\begin{equation}
h_\pm(t)=
\pm\frac{4a^{3/2}}{3\sqrt{3b}}t^{3/2}.
\end{equation}
These are exactly the $L^{\pm}$ lines defined in \cite{li2024thermodynamic}. By contrast, maximizing along isotherms gives $h=0$, that is exactly the single-line Ising-symmetric Widom construction \cite{xu2005relation}.

The Landau model makes the range of this family explicit. As $\zeta$ runs from $-\infty$ to $+\infty$, $y_+(\zeta)$, and hence $x_+=y_++y_+^3$, increases monotonically from $0$ to $\infty$. Since $y_-(\zeta)=-y_+(-\zeta)$, the width of the intermediate region is $\mathcal{W}=(a^{3/2}/\sqrt b)\,[x_+(\zeta)+x_+(-\zeta)]\,t^{3/2}$. In units of the normalized field $H$, the $L^\pm$ prescription gives $\mathcal{W}/t^{3/2}=8/(3\sqrt3)\simeq1.54$, whereas $\zeta=\pm1$ gives $\mathcal{W}/t^{3/2}\simeq33$, about $22$ times wider. Since $y_+(\zeta)$ is increasing and convex, $\mathcal{W}$ is an even, convex function of $\zeta$, minimal at the $L^\pm$ prescription within this family (see Sec.~\ref{sec:special} for narrower regions obtained from other protocols). Individual branches, however, are not bounded by the $L^\pm$ lines: for $\zeta>0$ the positive branch lies outside $L^+$ while the negative branch lies inside $L^-$, and vice versa for $\zeta<0$. In the van der Waals realization of Fig.~\ref{fig:vdw}, the liquid-side branch at $\zeta=-2$ and the gas-side branch at $\zeta=+2$ both lie outside the $L^\pm$ lines, while their partner branches lie inside.

The $L^\pm$ lines nevertheless admit a transparent physical characterization in this model: the condition $3y^2=1$ is equivalent to $at=3bm^2$, i.e., to the balance between the thermal and the field-induced contributions to the inverse susceptibility, $\chi^{-1}=at+3bm^2$. This reflects a well-defined feature of the Landau scaling function. This balance, however, locates the crossover between the thermal-dominated ($|x|\ll1$) and field-dominated ($|x|\gg1$) asymptotic regimes only up to a factor of order unity: equally natural balances between the two terms of the equation of state or of the Landau free energy give $x=2$ and $x=3\sqrt2$, respectively, compared with $x=4/(3\sqrt3)\simeq0.77$ for the $L^\pm$ lines.

The sensitivity to regular thermodynamic contributions discussed above can be demonstrated explicitly within the Landau model. Consider its equilibrium free energy $f_0(t,h)$, whose equation of state is $h=atm+bm^3$, and add the analytic contribution
\begin{equation}
    \widetilde f(t,h)
=
f_0(t,h)-\frac{\varepsilon}{2}th^2.
\end{equation}
This modification is analytic and preserves the $h\to-h$ symmetry of the free energy. It therefore leaves the critical singularity unchanged. Since the added term does not depend on $m$, the Landau equation of state $h=atm+bm^3$ for the minimizer $m$ is unaffected. The physical order parameter, however, is the derivative of the full equilibrium free energy, $M=-\partial_h\widetilde f=m+\varepsilon th$, so that $m$ and $M$ no longer coincide and the measured susceptibility becomes
\begin{equation}
    \widetilde\chi(t,h)
=\frac{\partial M}{\partial h}
=
\frac{1}{at+3bm^2}+\varepsilon t.
\end{equation}
Using the scaling variables introduced previously,
\begin{equation}
    m=\sqrt{\frac ab}\,t^{1/2}y,
\qquad
h=\frac{a^{3/2}}{\sqrt b}t^{3/2}(y+y^3),
\end{equation}
where $y$ is still defined through the Landau minimizer $m$ rather than through $M$, the condition for a maximum along a constant-$h$ path becomes
\begin{equation}
    \frac{3y^2-1}{(1+3y^2)^3}
=-a\varepsilon t^2.
\end{equation}
For $\varepsilon=0$, this reduces to the original result $y_\pm=\pm1/\sqrt3$. For a small analytic perturbation, the solutions become
\begin{equation}
    y_\pm=
\pm\frac1{\sqrt3}
\mp\frac{4a}{\sqrt3}\varepsilon t^2
+O(\varepsilon^2t^4).
\end{equation}
Consequently, the crossover lines are shifted to
\begin{equation}
    \widetilde h_L^\pm(t)
=
\pm\frac{4a^{3/2}}{3\sqrt{3b}}t^{3/2}
\mp\frac{8a^{5/2}}{\sqrt{3b}}
\varepsilon t^{7/2}
+O(\varepsilon^2t^{11/2}).
\end{equation}
The analytic background therefore modifies the locations of the crossover lines while leaving both the critical point and the leading $t^{3/2}$ scaling unchanged. The displacement becomes subleading near criticality but is finite at any nonzero temperature distance from the critical point.

This provides a particularly transparent example of how supercritical response maxima can depend on nonsingular thermodynamic contributions that have no effect on the underlying critical singularity.
\subsection{Van der Waals fluid}

We start from the standard van der Waals equation of state in reduced variables,
\begin{equation}
    P_r=\frac{8T_r}{3v_r-1}-\frac{3}{v_r^2},
\end{equation}
where $P_r=P/P_c$, $T_r=T/T_c$, and $v_r=v/v_c$ denote the reduced pressure, temperature, and specific volume, respectively. The critical point is located at $P_r=T_r=v_r=1$.

Introducing the deviations from the critical point,
\begin{equation}
    t\equiv T_r-1,\qquad
    u\equiv v_r-1,\qquad
    p\equiv P_r-1,
\end{equation}
we expand the equation of state near criticality, obtaining
\begin{equation}
    p=4t-6tu-\frac32u^3
    +\mathcal{O}(tu^2,u^4).
\end{equation}
We can then identify the effective ordering field
\begin{equation}
    h=4t-p,
    \label{hvdw}
\end{equation}
which measures the pressure displacement from the critical isochore, so that
\begin{equation}
    h=6tu+\frac32u^3+\cdots
\end{equation}
where the omitted terms are subleading in the critical scaling limit $u\sim t^{1/2}$. This has precisely the Landau equation-of-state structure. As is well established, the van der Waals fluid therefore belongs to the mean-field universality class, with
\begin{equation}
    \beta=\frac12,\qquad
    \delta=3,\qquad
    \Delta=\frac32.
\end{equation}
The correspondence with the Landau model is $m\leftrightarrow u$, $a=6$, and $b=3/2$. Consequently, the analysis of the previous subsection applies directly to van der Waals fluids sufficiently close to the critical point.

In particular, the generalized path construction yields the family of crossover lines
\begin{equation}
    P_r^\pm(t;\zeta)
    =1+4t-A_\pm(\zeta)t^{3/2}
    +\mathcal{O}(t^2),
\end{equation}

where $A_\pm(\zeta)$ are the corresponding scaling amplitudes in the reduced-pressure normalization. The choice $\zeta=0$ recovers the fixed-ordering-field prescription underlying the $L^\pm$ lines of Ref.~\cite{li2024thermodynamic},
\begin{equation}
    P_r^\pm(t)
    =1+4t\mp\frac{16}{\sqrt3}t^{3/2}
    +\mathcal{O}(t^2).
\end{equation}
Changing the path parameter modifies the amplitudes without altering the leading critical exponents. This provides a direct realization of the nonuniqueness argument in a liquid--gas system.

Owing to the simplicity of the van der Waals model, we can go beyond the near-critical expansion and examine the crossover lines using the full equation of state. To this end, we consider the volume susceptibility
\begin{equation}
    \chi\equiv
\left(\frac{\partial v_r}{\partial h}\right)_t,
    \label{sus}
\end{equation}
which can be computed exactly as
\begin{equation}
    \chi(T_r,v_r)=
    \left[
    \frac{24T_r}{(3v_r-1)^2}
    -\frac{6}{v_r^3}
    \right]^{-1}.
\end{equation}
Here, $T_r$ and $v_r$ are the full reduced variables, rather than their deviations from criticality. Unlike the preceding expressions, this result does not rely on any near-critical approximation.

\begin{figure}
    \centering
    \includegraphics[width=\linewidth]{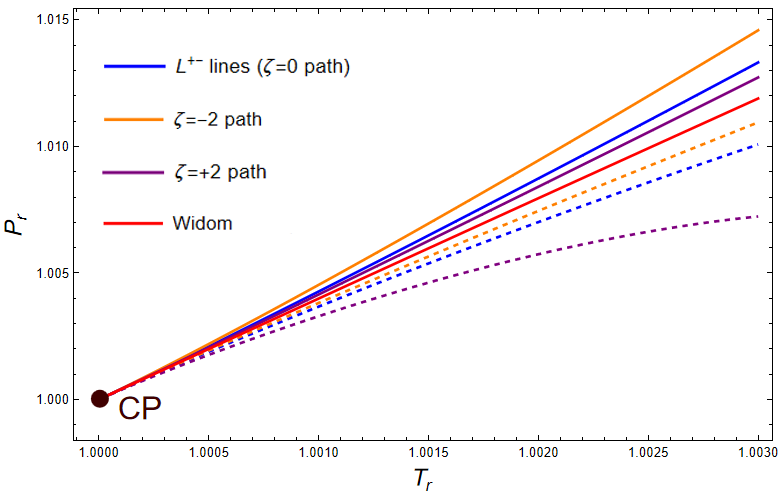}
   \caption{Nonuniqueness of thermodynamic crossover lines in the van der Waals model. For each choice of thermodynamic paths, the solid and dashed curves denote the two branches of maxima of the volume susceptibility. The blue curves are the $L^\pm$ lines of Ref.~\cite{li2024thermodynamic} ($\zeta=0$), while the purple and orange curves correspond to $\zeta=+2$ and $\zeta=-2$, respectively, with $\zeta$ defined with respect to the unnormalized field $h$ of Eq.~\eqref{hvdw}. The red curve is the Widom-type line obtained from isothermal maxima \cite{xu2005relation}, which is approached by one branch of each pair in the limit $|\zeta|\to\infty$, while the other branch moves away from it. All curves emanate from the critical point (CP) and exhibit the expected critical scaling sufficiently close to it. For each $\zeta\neq0$, one branch lies outside and the other inside the $L^\pm$ lines; e.g., the liquid-side branch at $\zeta=-2$ (orange solid) and the gas-side branch at $\zeta=+2$ (purple dashed) lie outside.}
   \label{fig:vdw}
\end{figure}

We again consider the continuous family of thermodynamic paths
\begin{equation}
    q=h-\zeta t^{3/2}=\mathrm{const.},
    \label{pathvdw}
\end{equation}
which, using the definition of the effective ordering field, can be written as
\begin{equation}
    P_r=1+4t-q-\zeta t^{3/2},
    \qquad t=T_r-1.
\end{equation}
Note that, here and in Fig.~\ref{fig:vdw}, $\zeta$ multiplies the unnormalized field $h$ of Eq.~\eqref{hvdw}, rather than the normalized field $H$ of Eq.~\eqref{fff}. Since $h=12H$ for the van der Waals fluid, the two parametrizations are related by $\zeta_h=12\,\zeta_H$.

Searching for stationary points of the susceptibility in Eq.~\eqref{sus} along these paths yields the exact condition
\begin{equation}
\begin{aligned}
&\frac{\partial P_r}{\partial v_r}
\frac{\partial^2P_r}{\partial v_r\partial T_r}\\
&\quad+
\frac{\partial^2P_r}{\partial v_r^2}
\left[
4-\frac32\zeta\sqrt{T_r-1}
-\frac{\partial P_r}{\partial T_r}
\right]=0.
\end{aligned}
\end{equation}
All derivatives are evaluated using the full van der Waals equation of state. Substituting them explicitly and simplifying, we obtain
\begin{widetext}
\begin{equation}
-8v_r\left[(3v_r-1)^2-4T_r v_r^3\right]
+\left[8T_r v_r^4-(3v_r-1)^3\right]
\left[12(v_r-1)-\frac{3}{2}\,\zeta\,(3v_r-1)\sqrt{T_r-1}\right]=0 .
\label{eq:vdw_exact_condition}
\end{equation}
\end{widetext}
This equation determines a continuous family of stationary loci in the $(T_r,v_r)$ plane, parameterized by $\zeta$. Retaining the physically stable branches corresponding to maxima of the susceptibility, and substituting the resulting solutions $v_r(T_r;\zeta)$ into the full van der Waals equation of state, we obtain the corresponding crossover lines $P_r(T_r;\zeta)$ in the reduced pressure--temperature phase diagram.

Figure~\ref{fig:vdw} shows representative examples of these lines, including the Widom-type prescription \cite{xu2005relation} and the $L^\pm$ construction \cite{li2024thermodynamic}. The remaining curves correspond to alternative choices of thermodynamic paths. These results illustrate that the location of response-function crossover lines depends explicitly on the operational prescription used to define them, even when the underlying equation of state and response function remain unchanged. Corrections to the leading-order amplitudes are largest on the gas side, where $|u|$ is largest, reflecting the asymmetry of the van der Waals equation of state. For instance, at $T_r=1.003$ the gas-side branch at $\zeta=+2$ lies at $P_r\simeq1.0072$, compared with $P_r\simeq1.0089$ at leading order.

\subsection{Black hole thermodynamics}

The preceding considerations also extend naturally to extended black-hole thermodynamics. In extended thermodynamic phase space, where the cosmological constant is interpreted as a pressure, charged anti-de Sitter (AdS) black holes exhibit a first-order small-to-large black-hole transition terminating at a critical point with van der Waals mean-field critical exponents \cite{Kubiznak:2012wp}. Recently, the $L^\pm$ construction of Ref.~\cite{li2024thermodynamic} has also been applied to these systems, leading to the identification of three distinct supercritical regimes \cite{doi:10.34133/research.1406}.

To illustrate that the same ambiguity arises in this context, we consider the four-dimensional Reissner--Nordstr\"om--AdS black hole at fixed electric charge $Q$. Its equation of state, expressed in reduced variables, reads
\begin{equation}
    P_r=\frac{8T_r}{3v_r}
    -\frac{2}{v_r^2}
    +\frac{1}{3v_r^4},
\end{equation}
where $P_r$, $T_r$, and $v_r$ again denote the reduced pressure, temperature, and specific volume, with the critical point located at $(T_r,P_r,v_r)=(1,1,1)$.

Introducing the deviations $t=T_r-1$, $u=v_r-1$, and $p=P_r-1$, the near-critical expansion gives
\begin{equation}
    p=\frac83t-\frac83tu
    -\frac43u^3+\cdots.
\end{equation}
Defining the effective ordering field
\begin{equation}
    h=\frac83t-p
\end{equation}
and retaining the leading critical scaling contributions, we obtain
\begin{equation}
    h=\frac83tu+\frac43u^3,
\end{equation}
which is again of the Landau form, now with $a=8/3$ and $b=4/3$.

Consequently, the generalized path construction developed above applies equally to this system. In particular, the fixed-ordering-field prescription yields
\begin{equation}
    h_\pm(t)
    =\pm\frac{32\sqrt6}{27}t^{3/2}
    +\mathcal{O}(t^2),
\end{equation}
corresponding to the two $L^\pm$ branches of Ref.~\cite{li2024thermodynamic} at leading order. Changing the thermodynamic paths generates infinitely many alternative crossover pairs satisfying the same critical scaling laws.

Thus, also in black-hole thermodynamics, the existence of the two $L^\pm$ lines does not establish their uniqueness or their interpretation as intrinsic boundaries separating distinct supercritical states. As in the van der Waals fluid, their location depends on the operational prescription used to identify the response-function maxima.

\section{Lee--Yang interpretation}
As elegantly established in the seminal works of Lee and Yang \cite{PhysRev.87.404,PhysRev.87.410}, phase transitions can be mathematically characterized through the complex zeros of the thermodynamic partition function. More recently, the Lee--Yang formalism has been extended to characterize supercritical crossover boundaries, including the Widom line \cite{PhysRevE.109.024118,xu2026thermodynamic}. Here, we revisit this approach and investigate whether it can resolve the ambiguities discussed above.

Consider the grand canonical partition function,
\begin{equation}
    \mathcal Z(T,z)=\sum_N Z_N(T)z^N,
\qquad z=e^{\mu/k_BT},
\end{equation}
where $z$ is the fugacity and $\mu$ the chemical potential. For a finite system, the partition function can be factorized in terms of its complex zeros $z_j$,
\begin{equation}
    \mathcal Z(T,z)=\prod_j\left(1-\frac{z}{z_j}\right).
\end{equation}
The thermodynamic potential is consequently related to these zeros through
\begin{equation}
    \Omega(T,z)=-k_BT\ln\mathcal Z(T,z).
\end{equation}
To establish the connection between the Lee-Yang zeros and the thermodynamic response maxima, it is convenient to express them in terms of a complex field $h$ conjugate to the order parameter. For a lattice gas, the fugacity is related to the Ising ordering field through $z\propto e^{2h/k_BT}$, so that zeros in $z$ map onto zeros $h_j$ in the complex field plane. In the following, we work directly with the latter. The susceptibility associated with a field coupled linearly to the order parameter takes the schematic form
\begin{equation}
    \chi(t,h)=\chi_{\mathrm{reg}}(t,h)
-\frac{k_BT}{V}\sum_j\frac{1}{(h-h_j)^2},
\end{equation}
where $\chi_{\mathrm{reg}}$ represents regular contributions. This expression makes clear that the response function on the real thermodynamic axis is influenced by the distribution of singularities in the complex plane.

To see this explicitly, consider first an isolated pair of complex-conjugate zeros,
\begin{equation}
    h_\pm=h_0\pm i\eta,
\end{equation}
where $\eta>0$ measures their distance from the real axis. Their contribution to the susceptibility is
\begin{equation}
    \chi_{\mathrm{pair}}(h)
=-\frac{k_BT}{V}
\left[
\frac{1}{(h-h_0-i\eta)^2}
+
\frac{1}{(h-h_0+i\eta)^2}
\right].
\end{equation}
Introducing $s=h-h_0$, we obtain
\begin{equation}
    \chi_{\mathrm{pair}}(s)=
\frac{2k_BT}{V}
\frac{\eta^2-s^2}{(\eta^2+s^2)^2}.
\end{equation}
This contribution exhibits a maximum at $s=0$, corresponding precisely to the real-axis projection of the conjugate pair,
\begin{equation}
    h_{\max}=h_0=\operatorname{Re}h_\pm.
\end{equation}
Moreover, its characteristic width is controlled by the imaginary part of these zeros, $\eta=|\operatorname{Im}h_\pm|$. Hence, as the zeros approach the real axis, their contribution to the response becomes increasingly pronounced and localized around their real projection.

This provides a simple explanation for the connection between complex Lee–Yang singularities and supercritical thermodynamic response maxima, as previously discussed in \cite{PhysRevE.109.024118}.

For a general system, however, the susceptibility receives contributions from all the complex zeros. To understand how they combine, consider a collection of conjugate pairs,
\begin{equation}
    h_j^\pm=h_{0,j}\pm i\eta_j,
\end{equation}
and assume that the relevant pairs lie sufficiently close to a common region of the real axis, such that $|h-h_{0,j}|\ll \eta_j$. Expanding their contributions gives
\begin{equation}
    \chi(h)\simeq\chi_{\mathrm{reg}}(h)
+\frac{2k_BT}{V}\sum_j
\left[
\frac{1}{\eta_j^2}
-\frac{3(h-h_{0,j})^2}{\eta_j^4}
+\cdots
\right].
\end{equation}

Neglecting the variation of the regular contribution, the condition  $\partial_h\chi=0$, that is the maximum of the corresponding susceptibility, yields
\begin{equation}
    \sum_j\frac{h_{\max}-h_{0,j}}{\eta_j^4}\simeq0,
\end{equation}
and therefore
\begin{equation}
    h_{\max}\simeq
\frac{\displaystyle\sum_j h_{0,j}/\eta_j^4}
{\displaystyle\sum_j 1/\eta_j^4}.
\label{simp}
\end{equation}

Therefore, the position of the response maximum is approximately a weighted average of the real projections of the complex zeros, with a stronger weight assigned to zeros closer to the real axis.

In particular, the weights scale as $\eta_j^{-4}$, implying that the nearest zeros can strongly influence the location of the maximum. If a single conjugate pair dominates, the maximum approaches its real-axis projection. However, we notice that this result is approximate. It assumes that the dominant zeros are sufficiently clustered, that their contributions can be expanded around the relevant maximum, and that regular contributions are negligible. More generally, the position of the maximum depends on the complete distribution of zeros and need not coincide exactly with the projection of any particular Lee–Yang edge.

The Lee--Yang formalism does offer something that response-function maxima lack. In the thermodynamic limit, complex singularities are intrinsic properties of the thermodynamic potential, independent of the paths used to explore the phase diagram. For an Ising-symmetric system above the critical point, the Lee--Yang zeros accumulate along the imaginary ordering-field axis, with edge singularities located at
\begin{equation}
h_{\rm YL}^{\pm}(t)=\pm i\,x_{\rm YL}\,t^\Delta,
\label{eq:YLedge}
\end{equation}
where $x_{\rm YL}>0$ denotes the edge location in the scaling variable $h/t^\Delta$, with its value depending on the normalization of the scaling fields. Being genuine singularities, their locations are unaffected by analytic additions to the equilibrium free energy. The Lee--Yang zeros are therefore insensitive to the second source of nonuniqueness discussed in Sec.~\ref{two}.

However, identifying intrinsic complex singularities does not automatically provide a unique way to construct crossover lines in the real thermodynamic plane. The edge in Eq.~\eqref{eq:YLedge} is a complex curve in the four-dimensional space of complex $(t,h)$, and a real line is obtained only after choosing a slice and a projection: which thermodynamic variable is continued to complex values, and how the resulting singularities are mapped onto physical thermodynamic coordinates. These prescriptions are not uniquely fixed by the singularities themselves.

Consider first the standard Lee--Yang construction, where the ordering field $h$ is continued to complex values at fixed real $t$. For ferromagnetic Ising systems, the Lee--Yang theorem guarantees that the zeros lie on the imaginary-field axis, so that ${\rm Re}\,h_j=0$. Their projections therefore all coincide with the symmetry axis $h=0$. Consistently, the weighted-average approximation of Eq.~\eqref{simp} identifies a maximum at $h=0$, corresponding to the conventional single-line Widom prescription in the Ising-symmetric scaling regime. This construction does not select two distinct crossover boundaries.

Consider instead the analytic continuation in temperature at fixed real ordering field, associated with complex-temperature singularities. Continuing the Lee--Yang edge relation $h=i\,x_{\rm YL}t^\Delta$ into the complex-temperature plane gives singularities at $\arg t=\pm\pi/(2\Delta)$, on the appropriate branches. Projecting these singularities onto the real-temperature axis using $t_{\rm proj}={\rm Re}\,t$, one obtains the pair of curves
\begin{equation}
h_\pm(t_{\rm proj})=
\pm\frac{x_{\rm YL}}{\cos^\Delta[\pi/(2\Delta)]}\,
t_{\rm proj}^\Delta .
\label{eq:Fisherlines}
\end{equation}
These curves exhibit precisely the same critical exponent as the $L^\pm$ lines, but generally have different amplitudes. For example, in the mean-field Landau model of Sec.~\ref{ggl}, expressed in terms of the normalized field $H$ of Eq.~\eqref{fff}, one has $x_{\rm YL}=2/(3\sqrt3)$ and $\Delta=3/2$, giving a projected amplitude $4\sqrt2/(3\sqrt3)$. By contrast, the $L^\pm$ prescription yields $4/(3\sqrt3)$. Choosing instead the modulus $|t|$ as the projected temperature distance gives yet another amplitude, $2/(3\sqrt3)$. A projection of this type was recently proposed as a definition of supercritical crossover lines for charged AdS black holes~\cite{fan2026emergent}.

Thus, different continuations and projections of the same complex singular structure generate distinct real-axis curves, all obeying the expected critical scaling, and none of them need coincide with the maxima of a particular response function. The Lee--Yang formalism therefore does not eliminate the nonuniqueness of supercritical crossover boundaries. Rather, it makes explicit the distinction between intrinsic complex singularities, which are fixed by the thermodynamics, and the operational procedures used to associate them with real-axis crossover lines, which are not. The existence of a characteristic complex singularity does not, by itself, establish a privileged partition of the supercritical region into distinct thermodynamic states.

\section{Is any crossover line special?}
\label{sec:special}

The preceding results establish that universal critical behavior permits infinitely many thermodynamic crossover lines. Even the number of distinct supercritical states is not fixed: depending on the protocol, the supercritical region divides into two states (liquid-like and gas-like) or three (with an intermediate region between the two branches). One may nevertheless ask
whether additional restrictions on the probing paths single out particular
constructions, and whether such restrictions carry physical significance.

Near an Ising-like critical point, the singular free energy is invariant
under $h\to-h$. For van der Waals-class fluids, this emergent symmetry renders the $L^\pm$ lines symmetric about the critical isochore in the scaling regime~\cite{cui2025emergent}. A family of probing paths is $Z_2$ covariant if reflection
maps each path into another path of the same family. The paths of
Eq.~\eqref{path} are not covariant for $\zeta\neq0$, since reflection maps
$\zeta$ into $-\zeta$. Within this family, only $\zeta=0$, corresponding
to the $L^\pm$ prescription, is covariant. This also explains why the
width is stationary at $\zeta=0$ within this particular family: for a
$Z_2$-symmetric response function, the width is even in $\zeta$.

Suppose, in addition, that the direction field of the probing paths is
analytic and nonvanishing at the critical point, a condition we call
regularity. On the symmetry axis $h=0$, covariance then permits only two
directions: tangent or transverse to that axis. For tangent paths,
$\zeta_{\rm eff}\to0$ in the scaling limit and the extrema approach the
$L^\pm$ loci whenever the chosen response function has two off-axis
maxima. For transverse paths, the extrema approach the central Widom-type
locus $h=0$. Thus, within this restricted class, the leading asymptotic
crossovers reduce to two different prescriptions rather than a unique one.

Neither restriction is imposed by thermodynamics. In particular, dropping
regularity while preserving covariance produces other symmetric crossover
pairs. The paths $h=q/t$ give $\zeta_{\rm eff}=-x/\Delta$ and, for the
Landau susceptibility, yield $x_\pm\simeq\pm0.36$ in units of the normalized
field $H$. Their width, $\mathcal W/t^{3/2}\simeq0.71$, is smaller than
$8/(3\sqrt3)\simeq1.54$ for the $L^\pm$ lines. Hence covariance alone
does not establish a minimum width or uniquely fix the crossover boundaries.

More fundamentally, the symmetry of the free energy does not oblige the
probing protocol to share that symmetry. Even imposing both covariance and
regularity leaves two inequivalent asymptotic prescriptions and does not
determine the response function or finite-distance locations. For real
liquid--gas criticality, the argument for these restrictions is weaker
still. The $Z_2$ symmetry emerges asymptotically in appropriately defined
scaling fields, which involve nonuniversal mixing of the physical
thermodynamic variables, while corrections to scaling introduce further
asymmetries~\cite{rehr1973revised,fisher2000yang,dj1t-tvw6}. A probing path that is
$Z_2$ covariant in scaling-field coordinates therefore need not correspond
to any distinguished experimental path in the physical $(T,P)$ plane.
Imposing covariance and regularity selects a mathematical class of
measurement protocols, not intrinsic thermodynamic boundaries.

Consequently, the apparent selection of Widom or $L^\pm$ lines follows
from additional assumptions about how the system is probed, not from an
independent physical principle. Without a protocol-independent criterion,
neither the number nor the location of supercritical crossover boundaries
is uniquely determined by the thermodynamics. The physical ambiguity remains.

\section{Discussion}
We have shown that supercritical crossover lines defined through thermodynamic criteria, particularly the loci of maxima of response functions, are inherently nonunique, since their locations depend explicitly on the thermodynamic paths along which the maxima are evaluated. This ambiguity adds to the already recognized dependence on the choice of response function itself. This path dependence is not the only source of ambiguity. Even for fixed paths and a fixed response function, the location of the maxima shifts under analytic modifications of the free energy that leave the critical singularity untouched, as shown in Sec.~\ref{two}. Finally, the Lee--Yang formalism, while providing an intrinsic and background-independent characterization of the complex singularities, does not remove the ambiguity: it reformulates it as a choice of complex continuation and of projection onto the real thermodynamic plane. Restricting the probing paths to regular, $Z_2$-covariant families selects the Widom and $L^\pm$ lines asymptotically, but only by an assumption on the protocol, and without fixing either the number of lines or their finite-distance locations.

The construction of Ref.~\cite{li2024thermodynamic} provides a well-defined and operationally transparent thermodynamic prescription, and the phenomena observed in its vicinity are genuine~\cite{3mxm-gqh5}. Our findings do not question the reproducibility of this construction or the physical significance of these phenomena. They show, however, that neither critical scaling nor symmetry identifies the $L^\pm$ lines as intrinsic boundaries between distinct thermodynamic states.

Although such crossovers can be experimentally accessible and reflect genuine changes in physical behavior between gas-like and liquid-like regimes~\cite{TORRESARENAS2026106969}, their precise locations do not necessarily represent intrinsic properties of the states they separate. In this sense, they cannot, on their own, be interpreted as genuine thermodynamic boundaries between distinct states of matter. The proposed existence of a third, intermediate regime in which liquid-like and gas-like characteristics become indistinguishable is similarly affected by this ambiguity. Indeed, we have demonstrated that the width of this region can be continuously varied by changing the thermodynamic paths used to define its boundaries, without modifying the underlying system, including to values smaller than that of the $L^\pm$ prescription. In particular, as shown explicitly for the Landau model, each branch of the admissible crossover lines individually sweeps the entire supercritical scaling region on its side, and the $L^\pm$ lines are neither its envelope nor its outer bound. What is universal is the existence of distinct asymptotic regimes of the scaling function, $|x|\ll1$ and $|x|\gg1$. The boundaries between them are crossovers whose location is fixed only up to a factor of order unity.

As Sec.~\ref{sec:special} shows, the Widom and $L^\pm$ prescriptions are singled out asymptotically only after restricting to regular, $Z_2$-covariant protocols. This restriction concerns how the system is probed, not the system itself, and even then it leaves two inequivalent answers. Physically, the location of these lines is therefore arbitrary: what is intrinsic is the continuous crossover between asymptotic regimes, not the lines drawn through it. More generally, the dependence of a proposed thermodynamic partition on the path used to explore the phase diagram reflects the absence of a genuine phase boundary beyond the critical point.

Alternative approaches based on dynamical or structural properties, such as the Frenkel line \cite{PhysRevE.85.031203}, are not directly subject to the same ambiguity associated with thermodynamic response maxima evaluated along different paths. However, they introduce their own operational criteria and need not terminate at the liquid--gas critical point. Such constructions may identify physically meaningful changes in microscopic dynamics without necessarily defining distinct thermodynamic states of matter.

Finally, in certain quantum spin models, dynamical singularities in quench dynamics have been shown to follow the same critical scaling as the $L^{\pm}$ lines \cite{hgwk-pzw2}. Whether the location of these singularities is likewise protocol dependent remains an open question.

\begin{acknowledgments}
We thank S.~Khrapak, W.~Li, X.~Li, Y.~Jin, L.~Li and Y.~Feng for useful discussions and comments on a preliminary version of this manuscript.
YX and MB acknowledge the support of the Foreign Young Scholars Research Fund Project (Grant No.~22Z033100604). MB is sponsored from the Yangyang Development Fund. We acknowledge the use of ChatGPT (OpenAI) and Claude (Anthropic) for assistance with language editing and analytical checks. The authors have independently verified the results and take full responsibility for the accuracy and content of this work.
\end{acknowledgments}

\end{document}